# User-Curated Image Collections: Modeling and Recommendation

Yuncheng Li
*University of Rochester*
*Department of Computer Science*
*Rochester, New York 14627, USA*
*Email: yli@cs.rochester.edu*

Yang Cong
*Chinese Academy of Sciences*
*Shenyang Institute of Automation*
*State Key Laboratory of Robotics*
*Email: congyang81@gmail.com*

Tao Mei
*Microsoft Research*
*Building 2, No. 5 Dan Ling Street*
*Haidian District, Beijing 100080, China*
*Email: tmei@microsoft.com*

Jiebo Luo
*University of Rochester*
*Department of Computer Science*
*Rochester, New York 14627, USA*
*Email: jiebo.luo@gmail.com*

*Abstract*—Most state-of-the-art image retrieval and recommendation systems predominantly focus on individual images. In contrast, socially curated image collections, condensing distinctive yet coherent images into one set, are largely overlooked by the research communities. In this paper, we aim to design a novel recommendation system that can provide users with image collections relevant to individual personal preferences and interests. To this end, two key issues need to be addressed, i.e., image collection modeling and similarity measurement. For image collection modeling, we consider each image collection as a whole in a group sparse reconstruction framework and extract concise collection descriptors given the pretrained dictionaries. We then consider image collection recommendation as a dynamic similarity measurement problem in response to user's clicked image set, and employ a metric learner to measure the similarity between the image collection and the clicked image set. As there is no previous work directly comparable to this study, we implement several competitive baselines and related methods for comparison. The evaluations on a large scale Pinterest data set have validated the effectiveness of our proposed methods for modeling and recommending image collections.

*Keywords*-Image Collection; Similarity Measure; Visual Features; Sparse Representation; Metric Learning

## I. INTRODUCTION

Recent years have witnessed the pervasiveness of social media in people's daily lives. Among the popular social activities, we are especially interested in image-sharing on such platforms as Flickr and Pinterest. In addition to share images, people often organize the individual images into "collections", e.g., the *photosets* on Flickr or *pinboards* on Pinterest. As a primary way to organize individual images, human curated collections exhibit many unique characteristics, e.g., highly relevant common topic, diversified yet coherent information with little redundancy. In fact, such image collections are drawing increasing attention on commercial social media platforms. For example, Pinterest generates revenue through image *boards* to help users explore the WWW,

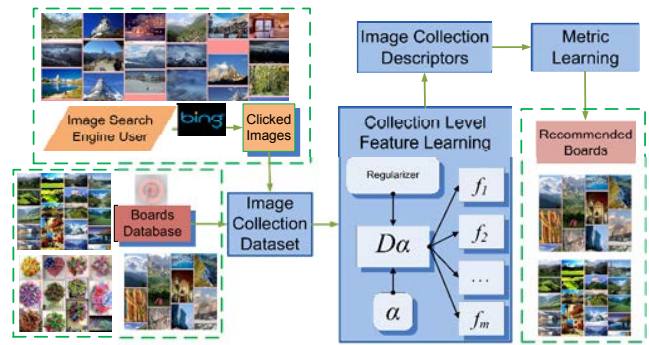

Figure 1: The diagram of our image collection modeling and recommendation framework, sharing the same notations as in Eqn. (1).

Facebook released a feature enabling users to collaboratively create and edit albums, and Bing returns image collections in response to image search queries. Even so, user curated image collections have been insufficiently studied in the research community.

In this paper, we intend to study modeling and recommendation of image collections, specifically for image search engine users with clicked images. For recommendation, we are taking a content based approach by comparing user's clicked images and image collection. An example application of this recommendation setting is a personalized search engine front page, where we can show high quality and up-to-date Pinterest boards according to the user's recent search history. There are two fundamental questions: 1) how to represent image collections, i.e., how to extract descriptors for a given image collection; 2) how to measure the similarity between user curated image collections and clicked images, in order to perform content based recommendation.

**Image Collection Modeling.** Although image collections

seem like simply groups of images, extracting a *discriminative* and *invariant* descriptor for them is a nontrivial task. We propose to model image collection based on sparse coding. In the image collection model, an image within a collection is treated as a "view" of the image collection. All views of an image collection can be jointly and sparsely reconstructed by a pre-learned dictionary, and the optimal reconstruction coefficients for these views are used as a compact descriptor for the image collection. We call this process *Collection Feature Learning (CFL)*, as illustrated in Fig. 1.

**Image Collection Recommendation.** The ultimate goal is to recommend image collections to image search engine users. Image search engine users are associated with a set of clicked images representing each personal preferences and interests, and the clicked images can also be seen as an image collection/set. While the cross domain nature of the problem disable traditional collaborative filter based recommendation, we recommend image collections by their similarity with clicked image set. Depending on the *CFL* for image collection modeling, we propose to use metric learning to adaptively learn a metric to evaluate the similarity, which is more robust and flexible than a fixed metric, e.g, Euclidean distance.

Fig. 1 shows the pipeline of the proposed recommendation system. Given an image search engine user, the system retrieves clicked images from the click through logs, and treat them as one image collection. In parallel, user curated image collections are extracted from a board database (image collections from Pinterest). Using *Collection Feature Learning*, descriptors are extracted from the entire image collection dataset, including Bing clicked image sets and Pinterest board dataset. Given the descriptors for the clicked image sets and Pinterest boards, and image search engine user's preference labels to Pinterest boards, a metric is learned to measure the similarity between clicked image sets and Pinterest boards. Next, the learned metric is used to rank and recommend Pinterest boards according to the similarity with a new user's clicked image sets. An example application of this recommendation setting is a personalized search engine front page, where we can show high quality and up-to-date Pinterest boards according to the user's recent search history.

This paper makes the following contributions:

- We define a new problem of discriminative image collection modeling. In comparison with previous research on image search that predominantly focused on ranking of individual images, we represent each image collection as a whole based on group sparse theory.
- We propose to use metric learning for adaptive recommendation and ranking. In order to facilitate more accurate matching between user interests and image collections across diverse categories, a similarity measurement is learned between the clicked image sets and a user curated image collection.
- We have built a new image collection dataset associated with user interests, by collecting large scale image collections from real-world online service platforms, i.e., Bing and Pinterest.

The organization of the remainder of this paper is as follows. In Sec. II, we review the related work. In Sec. III, we present our framework, including both image collection representation using sparse coding and image collection recommendation through metric learning. In Sec. IV, we introduce a new dataset for image collection research and report our initial experimental results and comparisons. Finally, Sec. V concludes the paper.

## II. RELATED WORK

In recent years, many social media recommendation systems have been proposed in the data mining community. Generally, two key issues need to be handled, object representation, i.e., how to extract various semantic information or descriptors; and object searching and ranking, which is in essence a similarity measurement problem. The word "object" we use here is a general concept, referring to all kinds of social media data, such as images, video and image collections in our case. In this paper, we focus on image collection recommendation.

### A. Image Sets

The most related works are on the modeling of image sets, such as Object Recognition with Image Sets (ORIS) [1], [2]. Based on a subspace assumption that an image set can be approximated by a low dimensional subspace, [3] proposed Grassmann Discriminant Analysis to find an optimal projection for image sets to achieve maximal discriminant power. Similarly, [1] proposed Manifold Discriminant Analysis that incorporates clustering techniques to further improve the subspace pursuit. While achieving success in object and face recognition, these works are different from our modeling of user curated image collection. First, the subspace assumption, held well with object recognition, are not suitable for the social image collections where users include more diverse content, which is empirically verified in the experiments section. Second, these methods are supervised learning, i.e., the class labels are required during subspace pursuit, so that they are not applicable in our recommendation scenario.

Moreover, there are other related works on image collection summarization. For example, [4] employed sparse reconstruction method to select representative and informative images from an image collection, and [5] proposed a clustering based method to effectively present image search engine results. Our image collection modeling is different from image collection summarizations, in that we are extracting discriminative descriptors for image collections (not representative images). It is worth noting that based on our observation, unlike image sets, there is little redundancy in user curated image collections precisely because during the curating process, the users carefully avoid redundancy to attract more views. For the same reason, this also demands a different collection representation model.

We assume that the clicked images are of interest. This is a reasonable assumption made by other systems that use clicked image sets to model user preferences. For example, [6] employed exploratory queries and the clicked images to improve personalized image search.

*B. Image Representation*

In this paper, we intend to represent an image collection, which is a group of highly related individual images, e.g., an image board from Pinterest [7]. For individual image representation, many important image representation methods are invented for various applications, including various Histogram-based features [8], [9], SIFT [10], GIST [11], Bag of Visual Words (BoVW) and spatial pyramid matching [12], kernel codebook [13], [14], Principal Component Analysis of Census Transform histograms (CENTRIST) [15], [16], the combination of CENTRIST and color cues [17], Color and Edge Directivity Descriptor (CEDD) [18], tiny image feature, Histograms of Oriented Gradients (HoG) feature [19], and so on. In contrast to the state-of-the-art individual images descriptors, we consider each image collection as a whole and use group sparsity to model the relationship among images in a collection.

Based on the basic image features, metric learning has been employed for information retrieval by adaptively learning the pairwise similarity, such as metric learning for medical image retrieval [20], and traditional single image retrieval [21]. Upon modeling the image collection, we further adopt metric learning to measure the distance between pairwise image collections.

### III. OUR APPROACH

In this section, we will explain image collection modeling and recommendation methods. The image collection modeling extracts concise collection descriptors based on a robust group sparse reconstruction formulation. With the collection descriptors, we propose an image collection recommendation system based on a metric learning scheme that measures search engine user's preferences on image collections.

*A. Social Image Collection Modeling*

Social image collections usually consist of several related images from 3rd party webpages. To our best knowledge, there is no method specifically designed for such an image collection. We intend to model a social image collection as a whole using the group sparsity model. We first extract several basic feature units from individual images. To eliminate the influence of redundant features and also improve the computation efficiency, we model image collections in a sparse coding framework based on $l_1$ and group sparse regularization with pre-learned concise dictionary. The details of our image collection modeling are explained as follows.

*1) Image Representation:* In this paper, we focus on using visual features, instead of user supplied textual information for several reasons. First, the textual information is often sparse and noisy on Pinterest. Second, we are interested in exploiting the rich visual information uniquely presented in a Pinterest board. It is conceivable that the sparse textual information can be integrated with rich visual information within this framework. We first extract several low-level and middle-level feature units from each individual image. Let $d$ denote the number of dimensions. The features we used include, GIST ($d = 320$) [11], CEDD ($d = 168$) [18], LAB Color Histogram ($d = 784$) [8], Tiny Image Feature ($d = 768$) and HoG ($d = 1984$) [19]. By concatenation, images are represented by 4024 dimensional features. Instead of fusing all kinds of features, for example image descriptions and collection titles.

*2) Dictionary Learning:* Even though we extract 5 individual feature units and wish they can complement each other to represent each image, there still exists redundant and noisy information. If we use them directly, it will not only increase the computation burden but also degrade the performance. Therefore, we first need to learn a concise dictionary from the training data. Previous works use unsupervised learning, e.g., Kmeans to cluster the training data into several clusters and aggregate these cluster centers together to construct the dictionary. In this paper, we adopt the dictionary learning model [22] based on $l_1$ minimization.

Suppose we extract the training data from image collections as $F \in \mathbb{R}^{d \times n} = \{f_1, f_2, \ldots, f_n\}, f_i \in \mathbb{R}^d$, where $n$ is the number of images, we aim to learn a concise dictionary $D \in \mathbb{R}^{d \times K}$ from $F$ with sparse coefficients denoted as $\alpha_i \in R^K$, which can be formulated as [22],

$$\min_{\alpha, D \in \Phi} \frac{1}{n} \sum_{i=1}^{n} \left( \frac{1}{2} \|f_i - D\alpha_i\|_2^2 + \lambda \|\alpha_i\|_1 \right), \quad (1)$$

where $\alpha = [\alpha_1, \ldots, \alpha_n]$ contains the sparse decomposition coefficients and $\lambda$ is the regularization parameter. $\Phi$ is the matrix verifying constraint, $\Phi = \{D \in \mathbb{R}^{d \times K} s.t. d_j^T d_j \leq 1\}, \forall j = 1, 2, \ldots, n$. The first term is sparse reconstruction error, which means the pursuit dictionary $D$ and coefficient $\alpha$ should reconstruct the training samples properly; the second term indicates that the coefficient $\alpha_i$ should be as sparse (zero dominant) as possible. The optimization of Eqn. (1) is based on the stochastic approximation, which is solved by optimizing $\alpha_i$ and $D$ alternatively.

To learn the diction efficiently, we train sub-dictionaries for each feature unit, using Eqn. (1) and concatenate these sub-dictionaries into a block diagonal matrix as $D = diag(D^i)$, which is depicted in Eqn. (2). Note that $\tilde{K} = 5K$ because there are five feature units in our case. The dictionary size $K$ is not sensitive according to our Experiments, so we fix it at 200 for all of our experiments.

$$D = \begin{pmatrix} D^1 & & \\ & \ddots & \\ & & D^5 \end{pmatrix} \in \mathbb{R}^{d \times \tilde{K}}, \tilde{K} = 5K \quad (2)$$

*3) Group Sparse Image Collection Modeling:* An intuitive idea to model a social image collection is averaging the feature vectors extracted from individual images in the collection. In our opinion, an image collection should be modeled holistically, because each image collection is created by a human curator with various information context, so there are some intrinsic relationships among the images. Another concern is to equip our model with the feature selection ability. Because the topic and content of images in an image collection are very diverse, we should select the most representative feature units to model the corresponding image collection, which will be more effective and robust.

Based on these two concerns and motivated by [17], we design our image collection descriptors:

$$\min_x \frac{1}{m} \left( \sum_{i=1}^{m} \frac{1}{2} \|f_i - Dx\|_h \right) + \lambda \Omega(x), \quad (3)$$

where $D \in \mathbb{R}^{d \times \tilde{K}}$ is the block diagonal dictionary as described above, $f_i \in \mathbb{R}^d$ is the extracted image features, $x \in \mathbb{R}^{\tilde{K}}$ is the estimated sparse coefficients, $m$ denotes the number of images in the image collection, and $\lambda$ is the regularization parameter. Using this formulation, the optimal joint reconstruction coefficients $x \in \mathbb{R}^{\tilde{K}}$ are used as a descriptor for the image collection.

In Eqn. (3), the first term shows that the estimated $x$ should represent all testing samples $f_i$ properly, and the second regularization term $\Omega(x)$ induces sparse $x$. There are many choices of the regularization term, such as $l_1$ and $l_g$. While $l_1$ produces sparse $x$ by selecting useful feature dimensions, $l_g$ enforces group sparsity by selecting basic image feature types, e.g., GIST and HoG. The group sparse regularization term $l_g$ in our case is defined first by breaking $x$ into groups $x^k$ corresponding to the feature types $f^k$ and merge them into a matrix $X = (x^1, x^2, x^3, x^4, x^5) \in R^{K \times 5}$, then $l_g(x) = \|X\|_{2,1}$. We emphasize that the $l_g$ group sparse regularization aims to reconstruct using small number of feature types, in contrast to using small number of feature dimensions with $l_1$ regularization. In the experiments, we empirically compare both methods in different image collection related tasks.

In Eqn. (3), the Huber estimator $\|\varepsilon\|_h$ is defined as in Eqn. (4), which is a smooth and robust estimator. Huber estimator tolerates outliers by the $l_1$ term that applys for large errors and penalize on inlier mistakes by the $l_2$ term that applys for small errors. The basic idea of Huber estimator is to Huber estimator is adopted here to eliminate the effects of inadvertently selected outlier images in the image collection, so that intrinsic features can be robustly extracted [23].

$$\|\varepsilon\|_h = \begin{cases} \frac{\varepsilon^2}{2\tau}, & \text{if } \|\varepsilon\| \leq \tau \\ |\varepsilon| - \tau/2, & \text{if } \|\varepsilon\| > \tau \end{cases} \quad (4)$$

*B. Image Collection Recommendation*

Now, given the feature descriptors of image collections, e.g., image board from Pinterest, we wish to measure the similarity between them and the Bing user clicked images, and rank them for recommendation accordingly. This is a distance measurement problem. For similarity measurement to search $K$ Nearest Neighboor (KNN), most previous works always choose and define various fixed local or global metrics, such as Euclidean metric, the Matusita metric, the Bhattacharyya coefficient, the Kullback-Leibler divergence. However, a fixed metric is unsuitable for our problem here. Because the topics of each image collection are so diverse, we cannot choose a suitable metric without enough prior knowledge. Even after we select many kinds of feature units for a better image collection representation, the challenges remain, as we cannot choose which feature unit makes more contributions in ranking. For example, while texture features may play little role for comparing cartoon image collections, it is very important for natural scene image collections.

Therefore, in this paper, we adopt metric learning [24] to adaptively learn the similarity in a more flexible way. Using the data mining techniques introduced in the experiments section (IV-A2), we can collect training data that signify user's preferences on image collections. For each user $u_i$, we obtain 1) clicked image set $c_i$, 2) interested boards $B_i^+$, and 3) disinterested boards $B_i^-$. Note that uppercase notation such as $B_i^+$ denotes a set of elements, while lowercase symbol such as $c_i$ denotes single element. The list of such associations $(u_i, c_i, B_i^+, B_i^-)$ are served as training data for the metric learning algorithm. Basically, we want to recommend $B_i^+$ not $B_i^-$ to a user who clicked $c_i$ on search engine. Therefore, the metric learning algorithm's goal is to find a metric that assign high similarity score to the pairs $S \triangleq \{(c_i, b) | b \in B_i^+\}$ and low similarity score to the pairs $D \triangleq \{(c_i, b) | b \in B_i^-\}$.

---

**Algorithm 1** Metric learning for recommendation

**Input:** Training data $(u_i, c_i, B_i^+, B_i^-), i = 1, 2, \ldots, U$
**Output:** Metric matrix $A$
1: Extract collection level features using collection level feature learning for all image collections $\{c_i, B_i^+, B_i^-\}$.
2: Construct similar pairs of collections $S$,
  $S \leftarrow \{(c_i, b) | b \in B_i^+\}$
3: Construct dissimilar pairs of collections $D$,
  $D \leftarrow \{(C_i, b) | b \in B_i^-\}$
4: Learn optimal $A$ using $S$ and $D$ according to Eqn. (6).

While there are many advanced metric learning algorithms developed for kNN classification, we adopt a straightforward method proposed in an early work [24] that best suits our "pairs only" case. The "pairs only" case means we can only define similar and dissimilar pairs $S$ and $D$, very different from the multi-class settings in kNN classification. Note that the paper is not to invent a new metric learning algorithm, but propose a framework to leverage existing metric learning methods to achieve image collection recommendation. For completeness, we introduce the metric learning method proposed in [24] in the following:

The first step is to parameterize the metric $d(x, y)$ between pairwise data instances as Mahalanobis distance,

$$d(x,y) = d_A(x,y) = \|x,y\|_A = \sqrt{(x-y)^T A(x-y)}, \quad (5)$$

where $A \in \mathbb{R}^{\tilde{K} \times \tilde{K}}$ is a positive semi-definite (PSD) matrix (the eigenvalues are non-negative), so that $d_A(x,y)$ satisfies the non-negativity and the triangle inequality constraints in metric definition.

Metric learning intends to learn a Mahalanobis distance thus the distance between similar pairwise samples is smaller than the distance calculated from dissimilar pairs. Next, the metric learning model is formulated as,

$$\begin{aligned} \max_A \quad & g(A) = \sum_{(x_i,x_j) \in D} \|x_i, x_j\|_A \\ s.t. \quad & f(A) = \sum_{(x_i,x_j) \in S} \|x_i, x_j\|_A^2 \leq 1 \\ & A \succeq 0, \end{aligned} \quad (6)$$

where $S$ and $D$ are the set of similar and dissimilar pairwise points, respectively. For model optimization, Eqn. (6) can be optimized using gradient ascent algorithm followed by iterative projection methods.

The training algorithm for metric learning is summarized in Alg. 1. Given the learned metric $A$, we can perform recommendation by ranking image collections $B$ in ascending order of its distance to the user's clicked image set $c$, i.e., $argsort_{b \in B} d_A(c, b)$.

## IV. EXPERIMENTS

In this section, we first report key statistics of our main database and introduce our data mining techniques to obtain training data. Next, we present experiments to verify the effectiveness of each module, i.e., image collection modeling and image collection recommendation.

### A. Main Database and Data Mining

The whole dataset is primarily composed of two parts, i.e., Bing click-through logs and Pinterest board dataset. Since our experimenting dataset was sampled from the main database, we demonstrate key statistics of the main database to show how our experiments on sampled datasets are closely related to realistic applications.

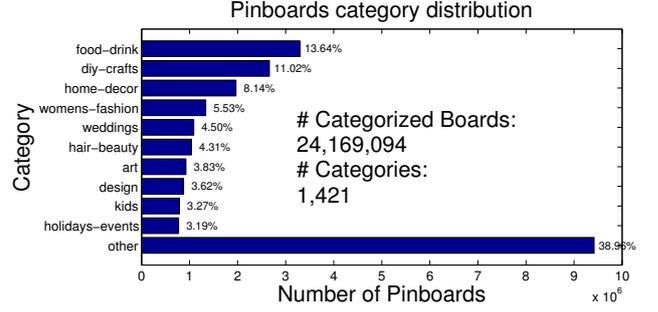

Figure 2: The distribution of the number of boards for top 10 categories, within all categorized boards from Pinterest database.

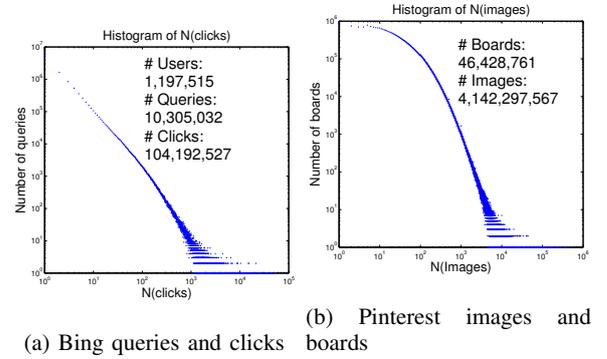

(a) Bing queries and clicks  (b) Pinterest images and boards

Figure 3: Dataset Statistics

*1) Main Database:* **Bing click-through logs** contain a list of user queries and the clicked images. We use $u_i$, $Q_i$ and $C_i$ to denote user, the associated queries and clicked image sets (one set for each query), respectively. By submitting queries and thus clicking images, users express their interests and preferences over images. In other words, user $u_i$ is interested in the clicked image set $C_i$. The full dataset is sampled from 5 days' Bing search logs using Bing API.

**Pinterest board database** contains a list of boards and the member images. Each board on Pinterest is a collection of images curated by users, which is one kind of image collection of this paper. Images within a board are pinned from 3rd party webpages by users to showcase specific topics. The number of member images varies significantly across all boards. On Pinterest, the information associated with a board including title, description, images and their titles, and category label assigned by board owners. Example category labels are *"Animals"* and *"MensFashion"*. Across all boards in the dataset, 52.06% of boards are categorized, within which the category distribution for top-10 categories are plotted in Fig. 2. The category distribution demonstrates also a heavy tailed distribution and most of the categorized boards are about people's daily life. The Pinterest board category labels are used to evaluate the proposed image collection modeling in a categorization task. This dataset is crawled from Pinterest using a number of seed webpages.

As shown in Fig. 3, the number of clicks per image query and the number of images in a Pinterest board exhibit power law characteristics, which indicates that the database is closely related to real cases and careful filtering is needed for further processing. Before the data mining step explained in the next section, we first apply filters to clean up the user and board database. We remove any user whose number of clicks is larger than the 95th percentile. We also remove any board whose number of pins is greater than the 95th percentile. It is a common practice to use 95th percentiles to remove data outliers [25].

*2) Data Mining User Preferences and Interests:* Our recommendation system works in a *cross domain* scenario, i.e., recommending Pinterest boards to Bing users. One of the biggest challenges for such a system is how to obtain the groundtruth for actual user preferences. One direction to solve this problem is to first conduct a user study by recruiting users to label their own preferences, or specialized judges to label real Bing user preferences, given each user's search log information. These methods suffer from subjectivity and scalability issues, thus usually work for some testing purposes. Another direction is to generate implicit feedbacks from real user activity logs using data mining techniques. We adopt the second method to mine Bing user's preference on Pinterest boards.

The mining process works as follows, (1) For user $u_i$, extract its *long query*, denoted by $q_i$ (2) Extract the clicks by user $u_i$ for $q_i$, denoted by $c_i$ (3) Extract Pinterest boards whose title are matched with this *long query*, denoted by $B_i^+$ (4) Insert $(q_i, c_i, B_i^+)$ into user $u_i$'s interested board pool (5) Sample the disinterested board list $B_i^-$ from unrelated queries, i.e, $\{B_j^+ | q_j \neq q_i\}$. (6) Output $(c_i, B_i^+, B_j^-)$ to the training data pool.

Because we are treating clicked image sets as a special image collection, the semantic consistency among the images should hold just as real image collections. This is why we are using clicked images from only one query for each training pair, because pooling all clicked images from a user will result in an image set that can not be treated as an image collection. However, this does not limit us to recommend image collections to users with multiple queries, because we can simply aggregate the recommendation results for individual clicked image sets.

The mining process leverage an observation that *long query* (longer than 3 words) carries little ambiguity, and thus directly reflect user's preference. Therefore, simply matching Pinterest board titles with long queries can accurately match user's preference. The text match is based on the length of longest continuous common subsequence (LCCS), which is widely used in information retrieval community to compute query document relevance, for example in the open source Sphinx search engine [26]. In order to obtain clean training data, we keep only the matches with highest LCCS scores. Note that this text based process works only for long queries, so for general queries, the proposed content based recommendation will play a more important role. In summary, using textual matching results for user's *long query*, we obtain the training data for content based models. For a new user who submitted a general query and clicked images for it, we recommend Pinterest boards by matching the clicked images using the content based models.

| Dataset | Items | Quantity |
|---|---|---|
| Image Collection Categorization | image collection categories | 14 |
| | average image collections per category | 100+ |
| | total image collections | 1600 |
| | total image | 30K+ |
| Image Collection Recommendation | *long queries* per user | 1 |
| | clicked images per user | 20 |
| | interested boards per user | 20 |
| | disinterested boards per user | 40 |
| | images per board | 20 |
| | total images | 36K+ |
| Query Dependent Metric Learning | query categories | 7 |
| | average query per category | 12 |
| | total number of queries | 84 |

Table I: Statistics of datasets used for evaluation.

Note that this specific mining process is used to evaluate our models, but the models will work seamlessly with other mining algorithms. We further randomly sample the mining results for evaluation purpose. Statistics for the final dataset for image collection recommendation are listed in Table I.

*3) Image Collection Categorization Dataset:* From top board categories, we select 14 categories with visual coherency, i.e., *CarsMotocycles, Travel, HolidaysEvents, Architecture, MensFashion, Celebrities, Humor, Animals, Weddings, WomensFashion, Photography, ScienceNature* and *Art*. From each category, we randomly sample 200 boards and for each board, 20 images are randomly sampled to model the board. After filtering invalid boards regarding these criteria, 1600 boards with 30K+ images in total are used to evaluate image collection modeling performance in a collection categorization task. Statistics for the datasets for both experiments are listed in Table I.

### B. Baselines and Evaluation Metrics

We conduct comparison experiments to empirically justify the proposed image collection models and metric learning based recommendation. We first explain the baselines and evaluation metrics.

*1) Image Collection Modeling:* In order to prove the effectiveness of the proposed image collection descriptors, we define and compare several variations, which are explained as below:

**Huber-L1** and **Huber-G:** Using $l_1$ (or $l_g$) as the regularization term $\Omega(x)$ in Eqn. (3) produces sparse (or *group* sparse) image collection descriptors. In addition, the Huber estimator is used to tolerate outliers in the image collection.

***Avg-L1*** and ***Avg-G:*** In order to validate the importance of Huber estimator, we construct a baseline by replacing the Huber estimator $\|\varepsilon\|_h$ with least square estimator $\|\varepsilon\|_{ols} = \varepsilon^1$ in Eqn. (3). With simple algebra derivations, using least square estimator, Eqn. (3) can be simplified to Eqn. (7), i.e., first compute the average feature vectors $\overline{f}$, then reconstruct the averaged feature vector using either $l_1$ minimization (*Avg-L1*) or $l_g$ minimization (*Avg-G*).

$$\min_x \frac{1}{2}\|\overline{f} - Dx\|_2^2 + \lambda\Omega(x), \quad (7)$$

***Raw-Avg:*** This is the most intuitive and straightforward baseline by simply averaging the image descriptor $f_i$ extracted from each individual image, i.e., $\overline{f}$. Another possibility is to match two sets by the most similar pair of images, which is in general not robust due to the apparent influence of outliers.

***MDA:*** In order to show the advantages over subspace based methods, we compare with the Manifold Discriminant Analysis (MDA) algorithm proposed in [1] as an example. Unlike the proposed collection modeling in this paper, the MDA algorithm does not extract descriptor for image sets.

*2) Image Collection Recommendation:* The image collection recommendation is based on metric learning by adaptively learning the similarity between user clicked image set and image collections. The metric learning model aims to learn optimal Mahalanobis metric $A$ under the defined objectives and we adopt several variations on the structure of $A$ for comparison, as explained below:

***Eucl:*** $A$ is fixed as an identity matrix. In this case, Eqn. (5) is equal to Euclidean distance.

***Diag:*** $A$ is constrained to be a diagonal matrix. In this setting, the metric is learned for each feature dimension independently. *Diag* has its limitations on the modeling of metric, e.g., not able to model the correlation between dimensions, but it enjoys the benefits of low dimensional parameter spaces, i.e., easy to optimize with stable performance.

***Full:*** In this setting, the only restriction on $A$ is the basic PSD constraint. *Full* model is able to model hidden dependencies between feature dimensions. However, it suffers huge parameters space $\sim O(d^2)$, prone to overfitting hence unstable performance.

Because the metric learning is built upon the *Collection Feature Learning*, various combinations of features type and metric types can be carried out. For example, *Huber-G-Diag* means combining *Huber-G* collection descriptor and *Diag* metric matrix.

The recommendation system is essentially a ranking engine, i.e., rank all collections in response to each user's clicked images and recommend top ranked collections to the user. Therefore, we adopt the ranking metric Mean Average Precision at K (MAP@K) as an evaluation metric for the recommendation system. MAP@K is designed for ranked list evaluation that takes both relevance and ranked position

| Avg-G | Avg-L1 | Huber-L1 | Huber-G | MDA | Raw-Avg |
|---|---|---|---|---|---|
| 38.50 | 52.48 | **54.60** | 42.04 | 18.77 | 7.84 |

Table II: Average accuracy (%) comparison for various collection descriptors in the image collection categorization task.

into consideration. In our scenario, the relevance is replaced by user's preference.

MAP@K is the mean of AP@K for all ranked lists, and AP@K is defined as follows similar to [27],

$$\text{AP@K} = \frac{\sum_{i=1}^{K} P(i) rel(i)}{K},$$

where $rel(i) \in \{0, 1\}$ is equal to 1, if the $i^{th}$ item in the list is relevant and vice versa; and $P(i) = \sum_{k=1}^{i} rel(k)/i$ is the average $rel(i)$ up to position $i$. Larger MAP@K means better ranking performance. By varying the ranking index $K$ in MAP@K, we can assess average ranking performance up to position $K$.

### C. Image Collection Categorization

In this section, we compare various image collection descriptors using the image collection categorization dataset. The classifier is based on traditional RBF kernel SVM[1] for descriptor based methods, e.g., *Huber-G*. On the other hand, for *MDA*, because it is a subspace based method and does not generate descriptors from image collections, we cannot use SVM. Therefore, we use the kNN classifier as the original paper did. For this multi-class categorization setting, we use the one-vs-all policy for training and the testing is depending on the held-out dataset by classifying each image collections into one of 14 categories. The parameters for RBF SVM and kNN are carefully tuned using 5-fold cross validation. For this multiclass classification problem, we employ average accuracy as evaluationi metric. Table. II shows the average accuracy for classification and several observations can be drawn from Table. II. First, Huber estimator(*Huber-\**) outperforms Least Square estimator (*Avg-\**). Social image collection, e.g., boards on Pinterest, may often contain several off-topic images. Therefore, Huber estimator is more robust against outliers than Least Square estimator that relies on the Gaussian reconstruction error assumption. Second, $l_1$ regularization outperforms group sparse regularization using $l_{2,1}$ norm. Group sparse regularization will discover dominate feature modalities by discarding irrelevant modalities, but it fails to find sparse solution within a group. This result shows that $l_1$ regularization is sufficient for sparse modeling for image collections in categorization task. Third, *Avg-Raw* fails to compete with any others, which is because it cannot properly model the relationship among images within an image collection. In addition, MDA performs poorly in this image collection categorization task, which is consistent to the observation that the subspace assumption required by manifold based method, e.g., MDA, does not hold with social image collections.

---
[1] http://www.csie.ntu.edu.tw/~cjlin/libsvm/

| Category | Long query examples |
|---|---|
| Art | "salvador dali famous painting" <br> "john singer sargent hagia sophia" |
| Photography | "ansel adam bird beach" <br> "older pictur rita hayworth" |
| Chart/Diagram | "number fraction third grade" <br> "pete cat color sheet" |
| Fashion | "elie saab couture 2012" <br> "hannah gown moniqu lhuillier" |
| Food | "angel food cake blueberry" <br> "rice krispi birthday cake" |
| Decorations | "christmas deco mesh wreath" <br> "black fire opal nevada" |
| Celebrities | "paul wesley nina dobrev" <br> "brigitt bardot life 1958" |
| Natural | "view alps zermatt switzerland" <br> "bern switzerland clock tower" |

Table III: Example of queries and query categories. Each row shows query examples in a query category. Generally, each query category contains 12 queries in average.

### D. Image Collection Recommendation

Using the user preference data mining method introduced in Sec. IV-A2, we extract training data that signify search engine user's preference on image collections. Specifically, for each user $u_i$, we extract the clicked image set $c_i$, interested Pinterest boards $B_i^+$ and disinterested Pinterest boards $B_i^-$. The list of all users in the dataset is first divided into training and testing set. The training set are fed into the metric learning algorithm, i.e., Alg. 1, we learn a Mahalanobis metric $A$ to measure the similarity score between clicked image set and Pinterest boards. For each user $u_j$ in the testing set, the learned metric $A$ is used to rank $B_j \triangleq B_j^+ \cup B_j^-$ in ascending order of its distance to the user's clicked image set $c_j$, i.e., $argsort_{b \in B_j} d_A(c_j, b)$. Given the ranked list of Pinterest boards and each board's known label (from $B_j^+$ or $B_j^-$), we can compute MAP@K by aggregating results for all testing users.

The above mentioned metric $A$ is learned globally for all users and queries. In order to demonstrate the flexibility of the proposed metric learning based recommendation framework, we break the dataset into several subsets according to the query category. Then train and test the metric learning model separately inside each category. In this way, semantic information is naturally incorporate into the metric learning algorithm. The query dependent recommendation pipeline works as follows. 1) A user issues a query and clicked several images for it; 2) we first determine the query category ($qc$) to choose the right metric $A_{qc}$; 3) then we rank collection database using $A_{qc}$. The statistics for the query categories are shown in Table I. The example queries for each category are shown in Table III. Note that the two experiments (categorization and recommendation) are independent because the query categories here and Table I have no direct relations with the native Pinterest board category labels in Section IV-C. In order to make robust recommendation, we keep only the query session with more than 20 clicked images, which amounts to about 10% image search queries on Bing.

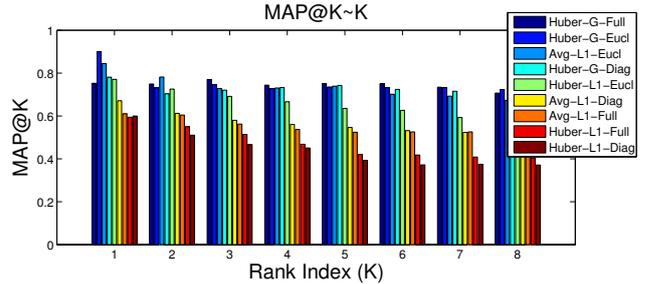

Figure 4: The image collection recommendation results, where the x axis and y axis are the ranking index and *MAP@K* scores, respectively.

|  | *Avg-L1* | | | *Huber-L1* | | | *Huber-G* | | |
|---|---|---|---|---|---|---|---|---|---|
|  | *Eucl* | *Full* | *Diag* | *Eucl* | *Full* | *Diag* | *Eucl* | *Full* | *Diag* |
| All | 0.59 | 0.55 | 0.52 | 0.61 | 0.52 | 0.40 | **0.78** | 0.77 | 0.77 |
| Art | **0.79** | 0.79 | 0.64 | 0.47 | 0.48 | 0.37 | 0.65 | 0.66 | 0.63 |
| Chart | **1.00** | 0.80 | **1.00** | **1.00** | 0.80 | 0.80 | **1.00** | **1.00** | **1.00** |
| Decorations | 0.72 | 0.72 | 0.68 | 0.67 | 0.25 | 0.69 | **0.88** | **0.88** | **0.88** |
| Fashion | **0.76** | 0.70 | 0.35 | 0.76 | 0.75 | 0.56 | 0.68 | 0.68 | 0.66 |
| Food | **1.00** | 0.00 | 0.20 | **1.00** | 0.04 | 0.30 | 0.45 | 0.76 | 0.64 |
| Outdoor | 0.71 | 0.52 | **1.00** | 0.52 | 0.40 | 0.00 | 0.64 | 0.54 | 0.54 |
| Celebrities | **0.65** | 0.60 | 0.46 | 0.53 | 0.49 | 0.37 | 0.54 | 0.47 | 0.55 |
| Photography | 0.42 | 0.04 | 0.07 | 0.17 | 0.07 | 0.05 | **1.00** | **1.00** | **1.00** |
| Mean MAP@5 | 0.74 | 0.52 | 0.55 | 0.64 | 0.42 | 0.39 | 0.74 | **0.75** | 0.74 |

Table IV: The MAP@5 performance comparison for the image collection recommendation. The top row refers to different collection descriptor, and the second row shows the variations of metric learning. The left column indicates the name of query categories (*All* means no query categorization is used) and each number is the value of MAP@5. The MAP@5 score for random ranking is 0.36, and all the baselines and proposed methods are significantly better than random guess.

We compare the performance of the various image collection representations and metric learning variants evaluated by MAP@5 as shown in Table IV, where the first row *Avg-L1*, *Huber-L1* and *Huber-G* are different collection descriptors, and the second row *Eucl*, *Diag* and *Full* are the metric learning variations. The MAP@K score averaged over all query categories at varying ranking index is illustrated in Fig. 4. Depending on this, we observe,

- *Huber-G* outperforms other image collection models. This result shows that group sparsity is useful in metric learning setting.
- The fixed metric *Eucl* performs better than learned

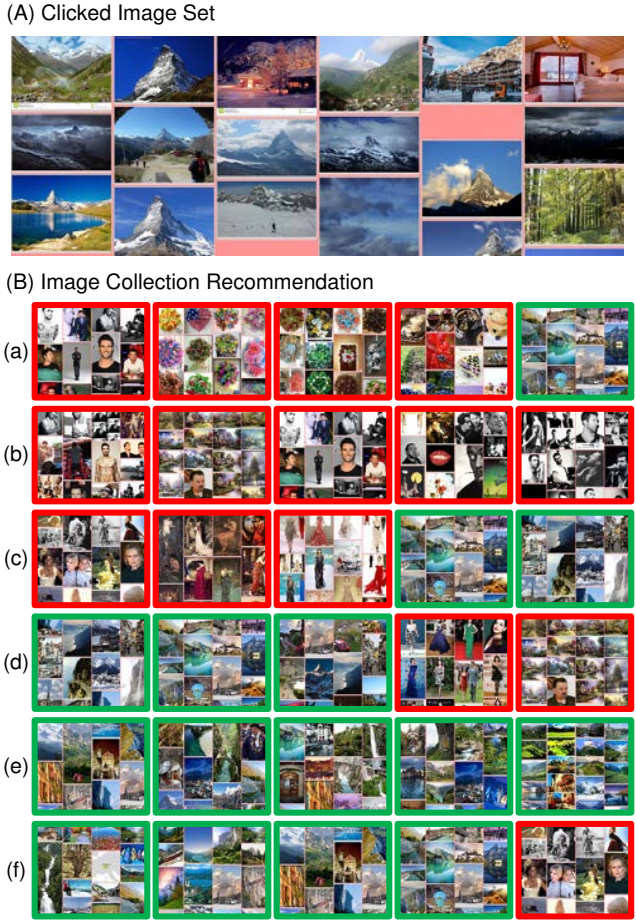

Figure 5: Example of image collection recommendation for a user who queried "View Alps Zermatt Switzerland". The top row (A) is clicked image set. Each row of (B) is the ranked top 5 image collection generated by (a) Avg-L1-Full, (b) Avg-L1-Diag, (c) Huber-L1-Full, (d) Huber-L1-Diag, (e) Huber-G-Full and (f) Huber-G-Diag, where the green and red rectangles indicate the interested and disinterested image collections, respectively.

metrics if no semantic information is incorporate into the recommendation, which shows the image collection modeling is already very effective for recommendation using global metric.
- When taking query category into consideration and learn query dependent metrics, perfect recommendation is achieved for some categories, which indicates the advantages of incorporating semantic information. For this scenario, metric learning performs better than fixed metrics *Eucl* in some categories, which indicates metric learning effectively discovered semantic specific structures in feature space.
- Simple *Eucl* metric can perform very well in most cases. This is because most noises has been removed by the sparse feature learning stage.
- Overall, *Full* metric is better. In some categories, *Diag* performs better than *Full*, consistent with the analysis that *Diag* enjoys the benefits of low dimensional parameter space.

A recommendation example, using various ranking protocols in response to a user's clicked images for the query "View Alps Zermatt Switzerland", is illustrated in Fig. 5. In general, the Huber estimator-based models perform better.

*E. Implementation Details and Runtime*

The optimization problems in image collection modeling and metric learning are handled by first order methods provided in the TFOCS package [28]. The free parameter $\lambda$'s are tuned independently to yield around 10% sparsity in the descriptors, which is align with the practice in [22]. The optimization routines are implemented in MATLAB. The collection modeling takes 2 seconds in average to extract descriptor for each collection, and the metric learning algorithm takes half an hour on average to learn a metric for each query category (in our experiment dataset scale).

## V. CONCLUSIONS

In this paper, instead of recommending individual images to users, we propose a new recommendation system that provides more meaningful and personalized image collections to users. A new dataset of image collections with implicit user preferences has been designed using data from Bing and Pinterest. Experiments and comparisons have demonstrated the effectiveness and robustness of the proposed modules, i.e., image collection models based on Huber estimator and sparse reconstruction, and image collection recommendation based on metric learning. Modeling and recommendation of socially curated image collections as holistic entities is a new topic in social media community, and should open the door to further research. Our future investigations include a user study to 1) compare our image collection recommendation with the traditional image recommendation, 2) compare the proposed collection descriptors with surrounding text in image collection recommendation, and 3) employ user specific metric learning to achieve ultimate personalization.